\documentclass[aps,pre,twocolumn,floatfix,showpacs,showkeys,letterpaper]{revtex4}

\usepackage{graphicx}
\usepackage{dcolumn}
\usepackage{natbib} 
\usepackage{bm,amsmath,amssymb,amsfonts,fancyhdr}

\newcommand{\pdiff}[3][]{\frac{\partial^{#1}
#2}{\partial #3^{#1}}}

\newcommand{\pder}[3][]{\frac{\partial^{#1} #2}{\partial #3^{#1}}}
\newcommand{\del}{\boldsymbol{\nabla}}

\newcommand{\divergence}[1]{\del\cdot{ \bf #1 }}

\newcommand{\laplacian}[1]{\nabla^2 #1}

\newcommand{\expec}[1]{\left< #1 \right>}

\newcommand{\bnabla}{\bm{\nabla}}
\newcommand{\bhat}[1]{\hat{\bm{#1}}}
\newcommand{\bvec}[1]{\vec{\bm{#1}}}
\newcommand{\bmr}[1]{\bm{\mathrm{#1}}}
\newcommand{\tor}{\mathrm{t}}
\newcommand{\Tor}{\mathrm{T}}
\newcommand{\pol}{\mathrm{s}}
\newcommand{\Pol}{\mathrm{S}}

\newcommand{\lm}[1]{#1_{\ell,m}}

\newcommand{\rmc}{Rm_{crit}}

\begin{document}


\title{Numerical simulations of current generation and dynamo excitation
  in a mechanically-forced, turbulent flow}
\author{R. A. Bayliss} 
\author{C. B. Forest }
\email{cbforest@wisc.edu}
\author{M. D. Nornberg } 
\author{E. Spence}
\author{P. W. Terry }

\affiliation{Department of Physics \\
University of Wisconsin---Madison\\
1150 University Ave.\\ Madison, WI 53706}

\date{\today}

\newlength{\figurewidth}
\setlength{\figurewidth}{\textwidth}

\newlength{\figurewidthtwo}
\setlength{\figurewidthtwo}{0.45\textwidth}

\begin{abstract}

The role of turbulence in current generation and self-excitation of
magnetic fields has been studied in the geometry of a mechanically-driven,
spherical dynamo experiment, using a three dimensional numerical
computation.  A simple impeller model drives a flow which can
generate a growing magnetic field, depending upon the magnetic Reynolds
number $Rm=\mu_0 \sigma V a$ and the fluid Reynolds number $Re=V a/ \nu$
of the flow. For $Re<420$ the flow is laminar and the 
dynamo transition is governed by a simple threshold
in $Rm>100$, above which a growing magnetic eigenmode is observed that is primarily
of a dipole field tranverse to axis of symmetry of the flow.
In saturation the Lorentz force slows the flow such that the magnetic 
eigenmode  becomes marginally stable.
For $Re>420$  and $Rm \sim 100$ the flow becomes turbulent and the dynamo 
eigenmode is suppressed.
The mechanism of 
suppression is due to a combination of a time varying large-scale field
and the presence of fluctuation driven currents (such as those
predicted by the mean-field theory) which effectively 
enhance the magnetic diffusivity.  For higher $Rm$ a dynamo reappears,
however  the structure of the magnetic field is often different from the
laminar dynamo; it is dominated by a dipolar magnetic field  
aligned with the axis of symmetry of the mean-flow which is apparently 
generated by fluctuation-driven currents.
The magnitude and structure of the fluctuation-driven currents has been
studied by applying a weak, axisymmetric seed magnetic field to 
laminar and turbulent flows. An Ohm's law analysis of the axisymmetric
currents allows the fluctuation-driven currents to be identified.
The magnetic fields generated by the fluctuations 
are significant: a dipole moment aligned with the symmetry
axis of the mean-flow is generated similar to those observed in the
experiment, and both toroidal and poloidal flux expulsion are observed.

\end{abstract}


\maketitle

\section{Introduction}
Astrophysical and geophysical magnetic fields are generated by complex
flows of plasmas or conducting fluids which convert gravitational
potential, thermal, and rotational energies into magnetic
energy \cite{moffatt78,childress95}. A comprehensive theory of the
magnetohydrodynamic dynamo is elusive since the generating mechanism can
vary dramatically from one system to another.  These variations arise
from differences in free energy sources, conductivity and viscosity of
the conducting media, and geometry.  Isolating and understanding
the mechanisms by which self-generation occurs and the
role of turbulence in the transition to a dynamo remain important
problems.

Dynamo action arises from the electromotive force (EMF) induced by the
movement of an electrically-conducting medium through a magnetic field.
This motional EMF generates a magnetic field which, depending on the 
details of the motion, can either amplify or
attenuate the initial magnetic field.  
If the induced field reinforces the initial magnetic field,
then the positive feedback leads to a growing magnetic field.
The source of energy for this dynamo is the kinetic energy of the
moving fluid.  The fluid may be driven by many different mechanisms such
as thermal convection in a rotating body for the case of the Earth, or
by impellers in liquid sodium dynamo experiments.

When the system is turbulent, the turbulence likely plays an 
important role in the dynamo onset
and the saturated state.  In the saturated state, the backreaction of the
self-generated magnetic field modifies the velocity field.
It is well known that in hydrodynamics, turbulence converts 
large-scale motions into smaller and smaller eddies, 
a process known as a turbulent cascade.
In magnetohydrodynamics (MHD),
fluid turbulence can fold a large-scale
magnetic field into  smaller structures \cite{kraichnan67}.
If the small-scale magnetic
fluctuations are helical, they can, on average, generate a net EMF
by interacting with the velocity field fluctuations and 
drive  large-scale currents. When the magnetic field of this fluctuation
driven current reinforces the original magnetic field self-excitation 
may be possible.
Thus the generation of
small-scale currents may explain observed large-scale magnetic
fields \cite{frisch75, pouquet78, alexakis05}.

Exact treatment of current generation in electrically-conducting fluids
requires solving the MHD equations governing the magnetic and velocity
fields:
\begin{eqnarray}
\pder{\bmr{B}}{t} &  = & 
\frac{1}{\mu_0\sigma}\nabla^2\bmr{B}+\bnabla\times \bmr{v}\times\bmr{B}
\label{eq:magind} \\
\rho \left[\pder{\bmr{v}}{t} + \left(\bmr{v} \cdot \bnabla\right) 
\bmr{v} \right] & = &  \bmr{J} \times \bmr{B} + \rho \nu \bnabla^2 \bmr{v}
-\bnabla {\it p} + \bmr{F}
\label{eq:motion} 
\end{eqnarray}
where $\rho$ is the density, $\sigma$ is the conductivity, $\nu$ is the
viscosity, and $p$ is the pressure.  ${\bf F}$ is a driving term
annotating the sundary sources of free energy in the flow.
Eqs.\ \ref{eq:magind} and \ref{eq:motion} are highly nonlinear and,
without limiting assumptions, are analytically intractable.
Early dynamo theory focused on solving only Eq.\ \ref{eq:magind} in the
kinematic limit where the linear magnetic field stability of
prescribed velocity fields was calculated to determine whether magnetic field
growth was possible \cite{bullard54,roberts72,gubbins73}.  
Due to advances in computing power during the
last decade, great progress has been made by performing numerical
simulations of dynamos, which simultaneously solve the non-linear
MHD equations (Eqs.~\ref{eq:magind} and \ref{eq:motion}).  
These studies break into two separate classes:
 global simulations which attempt to model geophysical or
astrophysical dynamos such as the Earth and the Sun \cite{glatzmaier84,glatzmaier95,glatzmaier02,
kageyama95,kageyama97,kuang97}, and simplified
models in which the geometry is simple enough 
to uniquely identify particular physical effects  \cite{meneguzzi81,cattaneo96,mueller03}.

The numerical simulations have been useful for studying magnetic field
generation, even though they are still far away from being able
to resolve the fluid turbulence of the actual
systems.  
In particular, the role of the magnetic
Prandtl number,  $Pm = Rm/Re$,
on threshold conditions for magnetic field growth is of
importance for understanding magnetic field generation in the Earth,
Sun, and in experiments. 
The linear self-excitation of the magnetic field is governed 
by the magnetic Reynolds number, $Rm=\mu_0 \sigma L V_0$, where
$\sigma$ is the molecular electrical conductivity, $L$ is a
characteristic size of the conducting region, and $V_0$ is the peak
speed.  Hydrodynamic turbulence is governed by the fluid Reynolds, number
 $Re=V_0 \ell/ \nu $, where
$\ell$ is the characteristic size of the flow.  
Simulations are capable of resolving the
modest $Rm$s needed to observe self-excitation, but not at the
very high values of $Re$ typical of low $Pm$ dynamos.
Recent studies 
in periodic boxes 
\cite{boldyrev04,schekochihin04} have focused on understanding the
generation of small-scale magnetic fields at low $Pm$, and 
simulations in cylindrical geometries with mean-flows have been performed \cite{ponty05} 
which show that the dynamo can be suppressed when turbulence is present.
The periodic box simulations are particularly good at modeling infinite,
homogeneous turbulence, though these conditions are rarely, if ever,
realized in actual astophysical or planetary contexts.
Little work has been done to understand the dependence of
large-scale magnetic field generation on $Pm$.

To address more realistic models of astrophysical turbulence, research
has turned to experiments.  Experiments at
Riga \cite{gailitis00,gailitis01,gailitis04} and
Karlsruhe \cite{stieglitz01,mueller04,mueller04a} use pumps to create
flows of liquid metal through helical pipes.  These experiments are
designed to be laminar {\sl kinematic dynamos}, i.e.\ the average
velocity field of the liquid metal is designed (through impeller and
pipe geometry) to produce a magnetic field instability.  The motivation
for using liquid metal in the Riga and Karlsruhe experiments is to allow
helical flows, yet the conduction and flow paths are not simply
connected.  Dynamos in simply-connected geometries where the flow is
unconstrained have yet to be demonstrated in an experiment.

The self-excitation threshold of the Riga and Karlsruhe experiments is
governed by the magnetic Reynolds number.  
For particular flow geometries, the kinematic theory predicts a
critical magnetic Reynolds number, $Rm_{crit}$, for self-excitation such
that a dynamo transition is observed when $Rm>Rm_{crit}$.  An important
result from the Riga and Karlsruhe experiments is that the measured
$Rm_{crit}$ at which the dynamo action occurs is essentially governed by
the mean velocity field.  Turbulence which was constrained by 
the characteristic size of the channel, $\ell$, 
apparently played little role.

The kinematic theory does not provide a hydrodynamically consistent treatment of 
the fluid turbulence, and in
simply-connected dynamo experiments the turbulent fluid motion will be
pronounced.  According to measurements in hydrodynamic experiments, the
turbulent velocity fluctuations scale linearly with the mean velocity
such that $\tilde{v} = C
\left<V\right>$.  Mean field theory \cite{krause80}
predicts that turbulence can modify the effective conductivity of the liquid
metal.  Random advection creates a turbulent or anomalous resistivity
governed by the spatial and temporal scales of the random flow.  A
reduction in conductivity due to turbulent fluctuations was observed at
low magnetic Reynolds number in liquid sodium \cite{reighard01}.  The
scaling of this turbulent resistivity is readily obtained by iterating
on the magnetic field in the nonlinearity of Eq.\ \ref{eq:magind}, and
looking at the term that depends on gradients of $\bmr{B}$.  For large
$Rm$ in a fluid with homogeneous, isotropic turbulence, the turbulent
resistivity is proportional to $\tilde{v} \ell_v$, and produces a
turbulent modification to the molecular conductivity,
\begin{equation}
\label{sigma_t} \sigma_T = \frac{\sigma}{ 1 +
C Rm \ell_{v}/L },
\end{equation} 
where $\ell_v$ is a characteristic eddy size (presumed to be some
fraction of $L$).  The turbulent
resistivity, as described above, operates even if there is no clear
scale separation between the mean flow and the turbulence,
 or if mean quantities are nonzero.  The turbulent
conductivity should be used for estimating the dynamo threshold: $Rm =
\mu_0 \sigma_T V_{0} L > Rm_{crit}$ results in a dynamo.  Thus, the
onset condition in a turbulent flow is governed by
\begin{equation}
\label{eq:rm_mod}
Rm > \frac{Rm_{crit}}{ 1 - C \ell_v Rm_{crit}/L}.
\end{equation}
Note that the potentially singular denominator imposes a requirement on
the effectiveness of a particular flow pattern for self-excitation;
dynamos will only occur if $Rm_{crit}<\frac{L}{C\ell_v}$.

The small $Pm$ of liquid metals implies large fluctuation levels and a
turbulent conductivity.  The influence of turbulent conductivity on
self-excitation enters through the dimensionless number $C
Rm_{crit}\ell_v/L $.  Through fluid constraints, the flow-dependent
parameters $C$, $\ell_v$ and $Rm_{crit}$ can be manipulated.  In the
Karlsruhe experiment \cite{mueller02}, for example, $\ell_v$ is set by
the pipe dimensions, rather than the device size hence $\ell_v/L$ can be
taken to be a fraction of the ratio of the pipe dimensions to the device
size.  An upper bound would be $\ell_v/L=0.06$.  We take $C< 0.1$, and
$Rm_{crit}\sim40$, hence $CRm_{crit}\ell_v/L<0.24$. We expect therefore
that dynamo onset would be governed mainly by laminar predictions, as
found experimentally.

Turbulence plays a much greater role in governing
self-excitation in geophysical and solar dynamos since there are no
boundaries to keep small-scale flow from influencing the conducting
region, and the values $Pm$ in the Earth's core
and in the convection zone of the Sun ($Pm\sim 10^{-5}$ to $10^{-6}$ and
$10^{-7}$ respectively) \cite{roberts00,cattaneo02}.  This is also true for 
several experiments now underway which investigate
magnetic field generation in more turbulent configurations \cite{petrelis03,peffley00}.
  One such experiment, at the University
of Wisconsin-Madison, uses two impellers in a $1$ m diameter spherical
vessel, to generate flows of liquid sodium with $\bigr<V\bigl>>\ 15 \mbox{ ms}^{-1}$.
These flows are predicted by laminar theory to be
dynamos \cite{oconnell00}. The Madison experiment is expected to achieve
$Rm>150$ which exceeds $Rm_{crit}$ by a factor of two.  Such experiments
have prompted a number of theoretical investigations into whether
magnetic field generation is possible for the small Prandtl numbers of
liquid metals in experiments without a mean
flow \cite{schekochihin04,boldyrev04}.  The Madison Dynamo Experiment uses a simple two vortex flow which,
according to a laminar kinematic theory, produces a transverse dipole magnetic
field.  The experiment presents a unique opportunity to test the
numerical models; the spherical geometry makes it particularly
well-suited to being simulated, and the magnetic fields 
 can be fully resolved, though
the fluid turbulence, cannot be fully resolved by simulation 
since $Re\sim 10^{7}$ in the experiment. 

In this paper, three-dimensional direct-numerical simulations are used to 
model the dynamics of the
experiment.  
The simulations are used to predict the behavior of the experiment
and give guidance on what role turbulence might have on current generation
and self-excitation.  Section \ref{sec:numerics} of the paper
describes the numerical model used for solving the MHD equations.
Section \ref{sec:lam_dynamos} describes results from $Pm\sim 1$
simulations where the flow is laminar and gives an overview of the large
scale flow which is linearly unstable to magnetic eigenmode growth.
Section \ref{sec:turb_dynamos} describes dynamos at lower $Pm$ where
the flow become turbulent.   Section \ref{sec:applied}
presents simulations of a uniform magnetic field applied to axis of
symmetry of the mean-flow in which turbulence generated currents are
investigated in subcritical flows.

\section{Numerical Model \label{sec:numerics} \protect}
The numerical model used in this paper solves the MHD equations 
in a spherical geometry,  resolving the velocity field at the origin, and
has a forcing term which mimics the impellers used in the experiment.
The code is designed to simulate the behavior
of a spherical liquid sodium experiment.
Sodium, at $100 \ ^{\circ} \rm C$, is an electrically conducting fluid
fully described by the imcompressible, resistive, viscous MHD equations. 
The code uses a spherical harmonic decomposition 
of the vector potential of the velocity and
magnetic fields in the $\theta$ and $\phi$ directions, and finite
difference representation in the radial direction.

The dimensionless equations
which govern fluid momentum, magnetic induction, and solenoidal field
constraints are:
\begin{eqnarray}
\pdiff{\bmr{v}}{t}+Rm_0(\bmr{v}\cdot\bnabla)\bmr{v} &= &
-Rm_0\bnabla P \label{eq:nd_navier} \\ &+ & Pm\nabla^2\bmr{v}+Rm_0\bmr{F}+Rm_0\bmr{J}\times\bmr{B},\nonumber 
 \\
\pdiff{\bmr{B}}{t}&=&
Rm_0\bnabla\times\bmr{v}\times\bmr{B}+\bnabla^2\bmr{B},
\label{eq:nd_induction} \\
 \bnabla\cdot\bmr{v} &  = & 0 \label{eq:solen} \\
 \bnabla\cdot\bmr{B} & = & 0.
\end{eqnarray}
In these equations, the time has been normalized to a characteristic
resistive diffusion time of $\tau_{\sigma}=\mu_0 \sigma a^2$ where $a$
is the radius of the sphere, and the
velocity has been normalized to a characteristic velocity $V_0$ so that
$ Rm_0 =V_0a\mu_0\sigma$.
For the actual experimental device, the radius of sphere is $a=0.53
\mbox{ m}$; $Rm_0 = 100$ 
corresponds to a characteristic speed of $V_0\approx 15 \mbox{ ms}^{-1}$.
The vector field $\bmr{F}$ is a stirring term of order 1 which 
models the impellers in the experiments.  In practice, the velocity field
resulting from the stirring term has a peak normalized velocity 
different from one.  This resulting velocity field 
is used to define the resulting
magnetic Reynolds numbers for a specific simulation, ie.\ $Rm =max(V) Rm_0$.
The relative importance of the magnetic and viscous
dissipation is expressed by $Pm=\nu\mu_0\sigma$ which for liquid sodium
is 10$^{-5}$; the fluid Reynolds number $Re$
is directly related to the magnetic Reynolds number by $Re=Rm/Pm$. 
In practice, the simulations have only been carried out for
$Pm>0.1$ which is sufficient to observed turbulence in the flows, but
four orders of magnitude larger than in the experiments.

Since the fluid is incompressible, the density evolution is unimportant
and the pressure equation need not be evolved.  
Other numerical representations of a spherical MHD system
solve for the pressure as a constraint on the flow 
\cite{quartapelle95}, especially in systems like stellar convection
zones where compressibility is part of the dynamics \cite{glatzmaier84}.
This simulation does not evolve the pressure explicitly; rather
it solves for the vorticity.  Taking the curl of Eq.\
\ref{eq:nd_navier}, the expression for the time evolution of the
vorticity is,
\begin{eqnarray}
\frac{\partial\bm{\omega}}{\partial t} \  &= &
Rm_0\nabla\times\bmr{v}\times \bm{\omega} + Rm_0
\nabla\times\bmr{J}\times\bmr{B}\label{eq:vorticity} \\ & + & Pm 
\nabla^2\bm{\omega}+Rm_0\nabla\times\bmr{F},\mbox{ and} \nonumber \\
\label{eq:inversion}\bmr{v} &= & (\nabla\times)^{-1}\bm{\omega} .
\end{eqnarray}

The spectral decomposition is that of Bullard and Gellman (BG), in which the
velocity field is described by a spherical harmonic expansion of
toroidal $\tor$ and poloidal $\pol$ functions \cite{bullard54},
\begin{equation}
\label{eq:BullardV}
\bmr{v}  = \bnabla\times \tor \bvec{r} + 
\bnabla\times\bnabla\times \pol\bvec{r},
\end{equation}
and the magnetic field is described similarly
\begin{equation}
\label{eq:BullardB}
\bmr{B}  =\bnabla\times \Tor \bvec{r} + \bnabla\times\bnabla\times \Pol \bvec{r},
\end{equation}
where $\pol,\ \tor,\ \Pol$ and $\Tor$ are complex, scalar functions of $r, \
\theta\ ,$ and $\phi$. This representation
automatically satisfies Eq.\ \ref{eq:solen}.  
To decompose Eqs.\ \ref{eq:BullardV}, \ref{eq:BullardB}, each scalar function 
is projected onto a spherical
harmonic basis set, normalized by
$\lm{N}=\sqrt{(2\ell+1)(\ell-m)!}/\sqrt{4\pi(\ell+m)!}$:
$\lm{Y}(\theta,\phi)=\lm{N} P_{\ell}^{m}(\cos{\theta})e^{im\phi}$.
$\lm{Y}$ is summed from $m=0, ... ,\ell$ and an extra factor of $\sqrt{2}$
in $\lm{N}$ for $m\ne 0$ since the function represents a real
field.  The result for the magnetic field is
\begin{align}
\Tor(r,\theta,\phi,t) & = \lm{N}\sum_{\ell=1}^{\infty}\sum_{m=0}^{m=\ell}
\lm{\Tor}(r,t)P_{\ell}^{m}(\cos{\theta})e^{im\phi} \\
\Pol(r,\theta,\phi,t) & = \lm{N}\sum_{\ell=1}^{\infty}\sum_{m=0}^{m=\ell}
\lm{\Pol}(r,t)P_{\ell}^{m}(\cos{\theta})e^{im\phi} 
\end{align}
and similarly for the flow scalars, $\tor$ and $\pol$.

One advantage of the BG representation is that
multiple curls, which appear with every poloidal component of the vector
fields, reduce to Laplacians.  The curl of a general solenoidal
vector-field, $\bmr{W}$ can also be represented by two scalar
functions of position, $e \mbox{ and } f$.  If
$\bmr{W}=\bnabla\times e \bvec{r} + \bnabla\times\bnabla\times f
\bvec{r}$, then clearly
\begin{equation}
\bnabla \times \bmr{W} = 
\bnabla\times (-\nabla^2 f \bvec{r}) + \bnabla\times\bnabla\times e \bvec{r}.
\end{equation}

To determine the discretized version of the vorticity equations, Eq.\
\ref{eq:vorticity} is expressed in terms of the toroidal-poloidal
representation:
\begin{equation}
\bm{\omega}=\bm{\omega_{\Pol}}+\bm{\omega_{\Tor}}=
\bnabla\times\bnabla\times \tor \bvec{r}+\bnabla\times (-\nabla^2 \pol)\bvec{r}.
\end{equation}
By substituting this form  into
Eq.\ \ref{eq:vorticity},  the need to determine boundary conditions on
the vorticity is eliminated and only boundary conditions on the velocity
field scalars are required.  The evolution equations for the flow
advance become
\begin{eqnarray}
\pdiff{\tor}{t}-Pm\nabla^2\tor & = &
Rm_0\left[G\right]_{\bmr{\Pol}}+\left[\bnabla\times\bmr{F}\right]_{\bmr{\Pol}}\\
\pdiff{\nabla^2\pol}{t}+
Pm\nabla^4\pol & = & Rm_0\left[G\right]_{\bmr{\Tor}}+
\left[\bnabla\times\bmr{F}\right]_{\bmr{\Tor}},\label{eq:bg_vort}
\end{eqnarray}
where $G$ signifies the sum of the advection and Lorentz forces.  The 
fourth-order derivative can be computed by consecutive Laplacian
operators.

The Crank-Nicolson method is used to advance the linear terms.  This
method implicitly averages the diffusive terms and computes a temporal
derivative accurate to second order.  The fluid advection term has a
hyperbolic character due to the propagation of inertial waves making it
advantageous to use an explicit advancement for nonlinear terms.  An
explicit second-order Adams-Bashforth predictor-corrector scheme is used
to advance the pseudospectral nonlinear terms.

The pseudospectral method computes a function in real space and then
decomposes it in spectral space.  Pseudospectral methods avoid the
complications of the full-spectral methods which rely on term-by-term
integrations of spectral components (such as in the Galerkin method) and
in general are much faster than full-spectral methods \cite{canuto88}.
The pseudospectral method has the disadvantage of introducing
discretization error through aliasing.  This error is addressed by
padding and truncating the spectrum \cite{canuto88}.

The radial derivatives in the diffusive terms are computed through
finite differencing on a nonuniform mesh.  The finite difference
coefficients for the $\nabla^2$ and $\nabla^4$ operators result in a
nonsymmetric band diagonal matrix.  The boundary conditions are folded
into the matrix defined by the implicit linear operators with
Gauss-Jordan reduction to ensure the matrix remains band-diagonal for
ease of inversion.  Using an optimized LU decomposition, the radial
evolution is solved independently for each spectral harmonic.  The
scalar fields are then converted to real space and the nonlinear cross
products are updated during predictor and corrector steps.

The temporal evolution loops over a spectral harmonic index, thus
individual boundary conditions for the respective harmonics are
separately applied.  The highest-order radial derivative in Eq.\
\ref{eq:bg_vort} is fourth order, requiring four boundary
conditions on the poloidal flow scalar.  Since the velocity must permit
a uniform flow through the origin, coordinate regularity implies
\begin{equation}
\begin{aligned}
\pol(r=0),\ \tor(r=0),\ \pder{\pol(r=0)}{r}  &= 0  &  \mbox{ for }  \ell \ \neq \  1 \\
\pol(r=0),\ \tor(r=0), \ \frac{\partial^2\pol(r=0)}{\partial r^2}
&= 0 & \mbox{ for } \ell \ = \ 1.
\end{aligned}
\end{equation}
For better numerical stability, the more stringent requirement $\pol, \
\tor \rightarrow r^{\ell}$ is applied to turbulent simulations.
The other boundary conditions are given by assumptions of a solid,
no-slip boundary.  For the poloidal flow,
\begin{align}
& \lm{\pol}(a)=0,\\ & \pdiff{\lm{\pol}}{r}\biggl|_{r=a}=0,
\end{align}
while for the toroidal flow
\begin{equation}
\lm{\tor}(a)=0.
\end{equation}

The discretization of the induction equation is straightforward in light
of the method presented for the flow.  Using the magnetic field given by
Eq.\ \ref{eq:BullardB}, the induction term in Eq.\
\ref{eq:nd_induction} is projected into toroidal and poloidal
components, grouping toroidal and poloidal contributions.  The
discretized expressions for the magnetic advance are:
\begin{eqnarray}
\pdiff{\Tor}{t}- \nabla^2\Tor & = &
Rm_0 N_T  \\
\pdiff{\Pol}{t}-
\nabla^2\Pol & = & Rm_0 N_S \label{eq:bg_mag}
\end{eqnarray}
where $N$ is the spectral transform of the inductive term in the BG
representation.  Coordinate regularity gives the conditions for the
magnetic scalar functions $\lm{\Pol}(r=0),\ \lm{\Tor}(r=0) = 0$.

The highest-order derivative of the magnetic advance is
$\mathcal{O}(r^2)$.  Given the conditions on the magnetic field at the
origin, a boundary condition on the magnetic field is needed at the
wall.  The outer surface of the Madison Dynamo Experiment is stainless
steel, modeled in the simulation as a solid insulating wall. The
remaining boundary conditions are solved by matching the poloidal
magnetic field to a vacuum field via a magnetostatic scalar potential,
and noting the toroidal field at the wall must be zero.  This implies
\begin{eqnarray}
\pder{\Pol_{\ell,m}}{r}+\frac{(\ell+1)}{a}\Pol_{\ell,m}&=&\frac{2\ell+1}{\ell+1}C_{\ell,m}\label{eq:Bpbc}, \\
\lm{\Tor}(a) & = & 0  .
\end{eqnarray}
In Eq.\ \ref{eq:Bpbc}, $\lm{C}$=0 if there are no currents in the
surrounding medium, but can also be finite to represent a magnetic field
applied by external sources.

The timestepping, while unconditionally stable for the diffusive
problem, is advectively-limited by an empirically-determined temporal
resolution requirement of $\Delta t \leq 5 (\Delta x)^2$ for a given
spatial resolution.  The spectral transform is the most
computationally-intensive portion of the code requiring roughly 80\% of
the CPU time.  The upper bound on the spatial resolution is: $
N_{\theta} \sim 64, \ N_{\phi} \sim 128, \ N_r \sim 400$ which gives,
with dealiasing, $\ell_{MAX}=42$, or nearly 1000 modes.

\begin{figure*}
\includegraphics[width=\figurewidth]{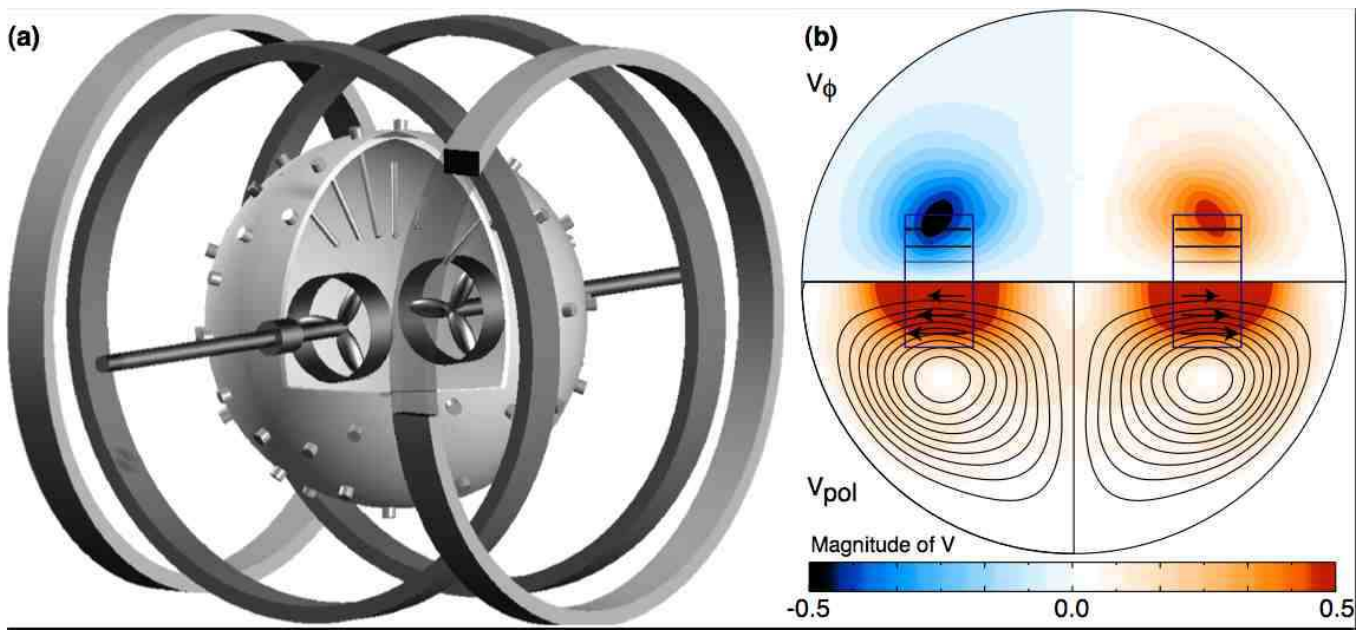}
\caption[A schematic of the Madison Dynamo Experiment and 
contours of simulated flow]{(a) A schematic of the Madison Dynamo
Experiment. The sphere is $1 \mbox{ m}$ in diameter and filled with
$105$--$110\ ^\circ$C liquid sodium.  High speed flows are created by
two counter rotating impellers. Two sets of coils, one coaxial and one
transverse to the drive shafts, are used to apply various magnetic
fields for probing the experiment.  (b) Contours of the toroidal
velocity $v_\phi$ and contours of the poloidal flow stream function 
$\Phi$ where $\bmr{v}_{pol} = \nabla \Phi \times \nabla \phi$, of the
axisymmetric double vortex flow generated by the impeller model. The
region of forcing is shown schematically along the symmetry axis.  }
\label{fig:1}
\end{figure*}

A forcing term, zero everywhere except at the location of the impellers in the
experiment, drives the flow. The forcing term for the impeller model is
\begin{eqnarray}
F_\phi(r,z) & = & \rho^2\sin{(\pi \rho b) }+\delta \\
F_Z(r,z) & = & \epsilon \sin{(\pi \rho c)}+\gamma. 
\label{eq:epsilon}
\end{eqnarray}
The axial coordinate, $z$,  and the cylindrical radius, $\rho$, are restricted to 
the region
 $0.25 a < \left| z \right| < 0.55 a$ and $ \rho < 0.3\, a$. 
The impeller pitch, $\epsilon$, 
changes the ratio of toroidal ($F_{\phi}$) to poloidal ($F_Z$)
force.  The constants $\delta$ and $\gamma$ control the axial force, and
in this article are zero in all but the applied-field runs where stronger
axial forcing is useful. The signs of $F_{\phi}$, $F_Z$ are positive for $z>0$ and
with $F_{\phi}$ negative for $z<0$ creating the counter-rotation between the flow cells.
$\bmr{F}$ is constant, which allows the input impeller power
$\bmr{F}\cdot\bmr{v}$ to vary. The region of the impellers and an
example of the resulting flow are shown in Fig.\ \ref{fig:1}(b).
These flows are topologically-similar to the ad hoc flows in several
kinematic dynamo studies \cite{dudley89,holme96,gubbins73},
but differ in that they are hydrodynamically consistent solutions to the momentum equation.

\section{Laminar Dynamos \label{sec:lam_dynamos}}
\begin{figure*}
\includegraphics[width=\figurewidth]{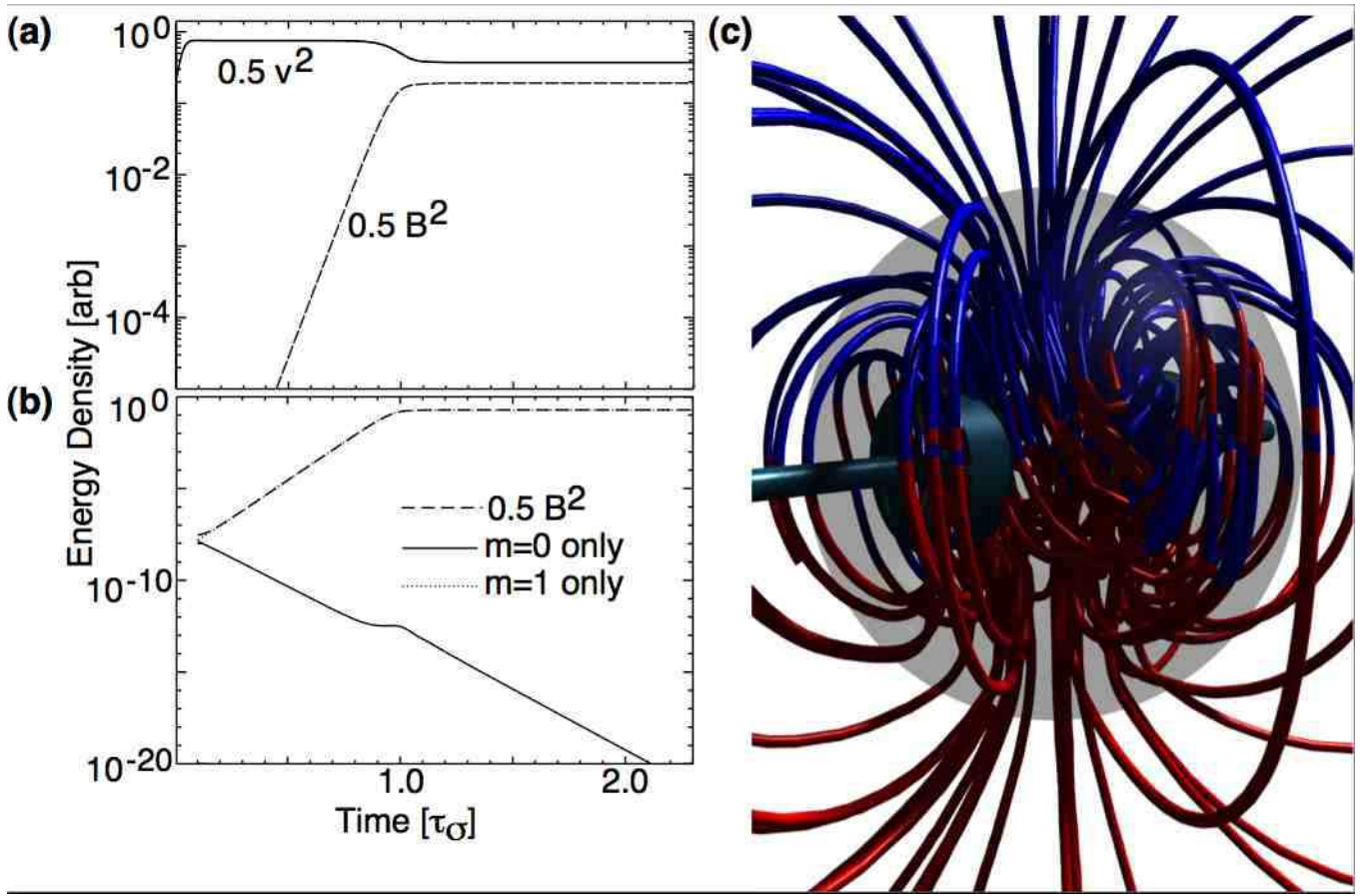}
\caption[Energy density and modal energy of a laminar dynamo]
{ (a) The kinetic and magnetic energy densities shown versus
time with $Rm=$159 and  $Pm=1$. 
The time is in units of the resistive diffusion time
$\tau_{\sigma}=\mu_0 \sigma a^2$. (b) The contributions to the total magnetic
energy density from the $m=1$ transverse dipole and the axisymmetric 
$m=0$ modes. (c) Magnetic field lines of a saturated dynamo state for a laminar flow with $Rm=150$.}
\label{fig:2} 
\end{figure*}

The impellor model described above predicts dynamo 
action for sufficiently-strong forcing.  For the particular case of
$Pm\sim 1$, a laminar flow results, as shown in Fig.\
\ref{fig:2}(a).  Starting from a stationary liquid metal, the
evolution is observed to go through several phases.  Initially, the
kinetic energy of the flow increases, as does the 
maximum speed ($Rm$) of the flow.  The resulting $Rm$  is
above the critical value at which dynamo
action is expected from kinematic theory.  The magnetic field energy
then increases exponentially with time. The measured growth rate agrees
with the growth rate $\lambda$ predicted by a kinematic eigenvalue code using the
generated velocity fields and solving  Eq.\ \ref{eq:nd_induction} for solutions
of the form $B\propto B_0 \exp^{\lambda t}$.
  After this linear-growth phase, a
backreaction of the magnetic field on the flow is observed which leads to
a saturation of the magnetic field.  In this saturated state, the
generated magnetic field is predominantly a dipole oriented transverse
to the symmetry axis, as seen  in Fig.\ \ref{fig:2}(c).  The $m=1$ 
equatorially dominant structure of the dynamo 
(shown in Fig.\ \ref{fig:2}(b)) is
consistent with kinematic analysis.

\begin{figure}
\includegraphics[width=\figurewidthtwo]{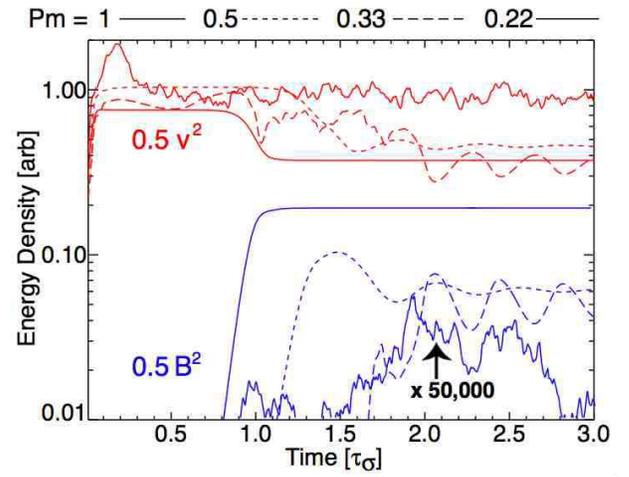}
\caption[Change in dynamo-saturation behavior with magnetic Prandtl number]
{ The magnetic and kinetic energy densities for runs with fixed $Rm$
($Rm=165\pm3\%$) but different $Pm$ versus time in
$\tau_{\sigma}$.  Note that $Pm=0.33$ shows a relaminarization of
turbulent flow while $Pm=0.22$ is barely amplifying the initial noise
and is shown multiplied by 50,000.}
\label{fig:5}
\end{figure}

The orientation of the generated dipole is not constrained by geometry
and is observed to vary between simulations.  When the saturation state
is oscillating, (or damped with oscillations as shown in the $Pm=0.5$
case in Fig.\ \ref{fig:5}) the dipole drifts around the
equator and also undergoes $180^\circ$ reversals.

\begin{figure}
\includegraphics[width=\figurewidthtwo]{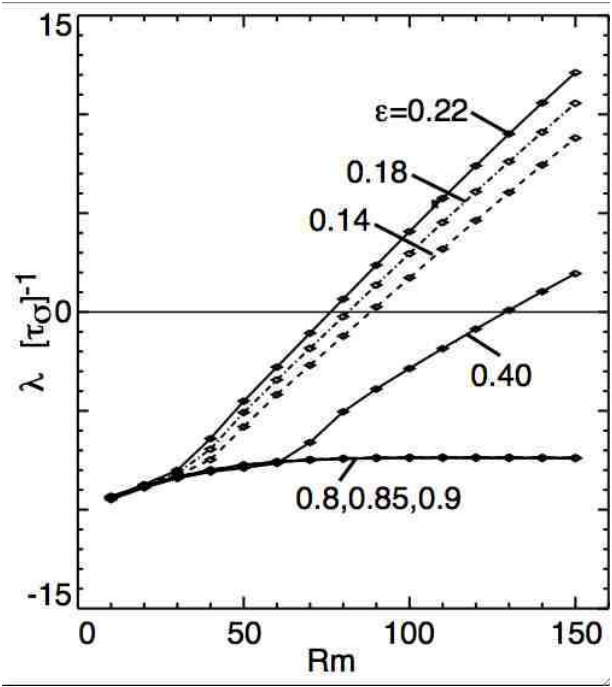}
\caption[Kinematic growth rates of simulated impeller pitches]
{The dependance of the linear growth rate of the least-damped magnetic
eigenmode on impeller pitch $\epsilon$.  The transition from damped to
growing ($\lambda=0$ point) defines the critical magnetic Reynolds
number; $Rm<Rm_{crit}$ to a growing magnetic eigenmode.}
\label{fig:3}
\end{figure}

Self-excitation depends on the shape of the flow as well as the
magnitude of $Rm$.  An ideal ratio of poloidal to toroidal forcing exists
(parameterized by $\epsilon$ in Eq.\ \ref{eq:epsilon}) for which the
critical magnetic Reynolds number is minimized as seen in Fig.\
\ref{fig:3}.  Minimizing $Rm_{crit}$ makes the flow easier to
attain experimentally. This optimal ratio can be understood from a
simple frozen flux model describing the stretch-twist-fold cycle of the
dynamo (see Ref.\ \cite{nornberg06}).  If the toroidal rotation is either too fast
or too slow relative to the poloidal flow, the advected field is not
folded back on to the initial field.

\begin{figure}
\includegraphics[width=\figurewidthtwo]{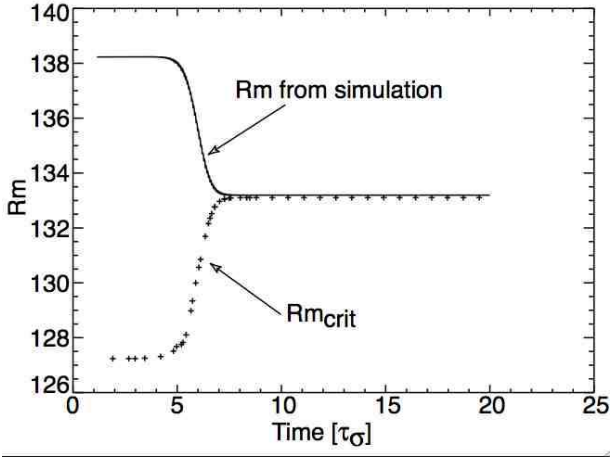}
\caption [The change in $\rmc$ during saturation of a laminar dynamo]
{ $Rm$ and $ Rm_{crit} $ evolution during the saturation phase of a
laminar dynamo.  $Rm_{crit}$ is calculated from linear stability for
each instantaneous velocity field during the simulation.  In saturation, $Rm=Rm_{crit}
$. }
\label{fig:4}
\end{figure}

For laminar flows, the backreaction is the result of two effects.
First, an axisymmetric component of the Lorentz force is generated by
the dynamo, slowing the flow and reducing $Rm$.  Second, the flow
geometry is changed such that the value of $Rm_{crit}$ is increased.  In
saturation the growth rate is decreased to zero, as the confluence of
$Rm$ and $Rm_{crit}$ in Fig.\ \ref{fig:4} shows.

\begin{figure}
\includegraphics[width=\figurewidthtwo]{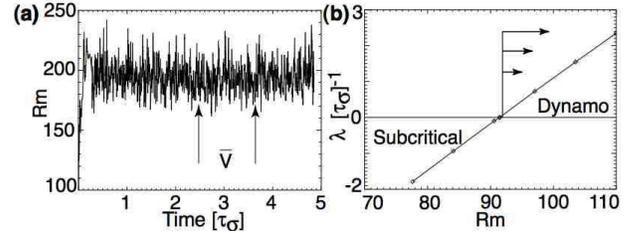}
\caption[Time behavior of Rm and kinematic analysis
of the mean flow in a turbulent run] {(a) Rm as a function of time.
Taking a series of flows over the range shown the mean flow is
calculated giving $Rm=193$ and $Re=863$.  (b) The kinematic analysis of
the average flow, where only Eq.\ \ref{eq:nd_induction} is evolved, and
$Rm$ is varied, to determine $Rm_{crit}\ = \ 91.5$.}
\label{fig:6}
\end{figure}

\begin{figure}
\includegraphics[width=\figurewidthtwo]{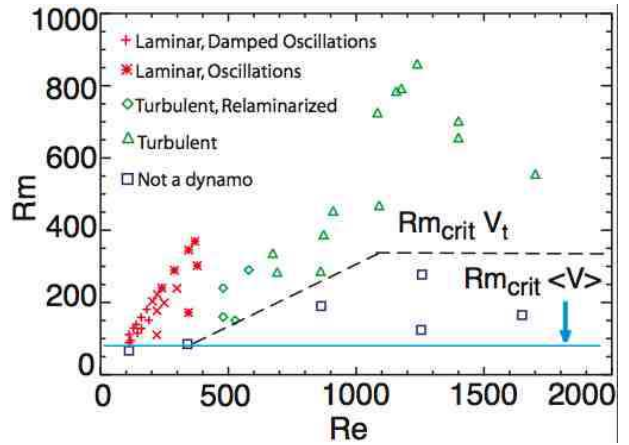}
\caption[$Re-Rm$ phase diagram.]
{ $Re-Rm$ phase diagram. A number of simulations whose hydrodynamic and
final saturated states are documented in Fig.\ \ref{fig:5}.
$Rm_{crit}$ for the mean flow $\left< {\bf V}\right>$ is essentially
independent of $Re$, while the effective dynamo threshold grows with
$Re$. The dashed line shows the qualitative behavior of the dynamo
threshold in turbulent flows ($V_t$).}
\label{fig:7}
\end{figure}

\begin{figure}
\includegraphics[width=\figurewidthtwo]{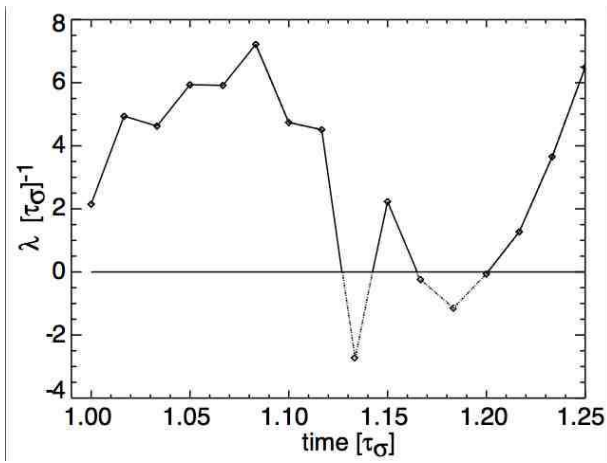}
\caption[Growth rate of the dominant eigenmode from a sequence
of turbulent flows] {The growth rate of the dominant eigenmode
calculated for a time-series of flow profiles.\label{fig:8}}
\end{figure}

\begin{figure*}
\includegraphics[width=\figurewidth]{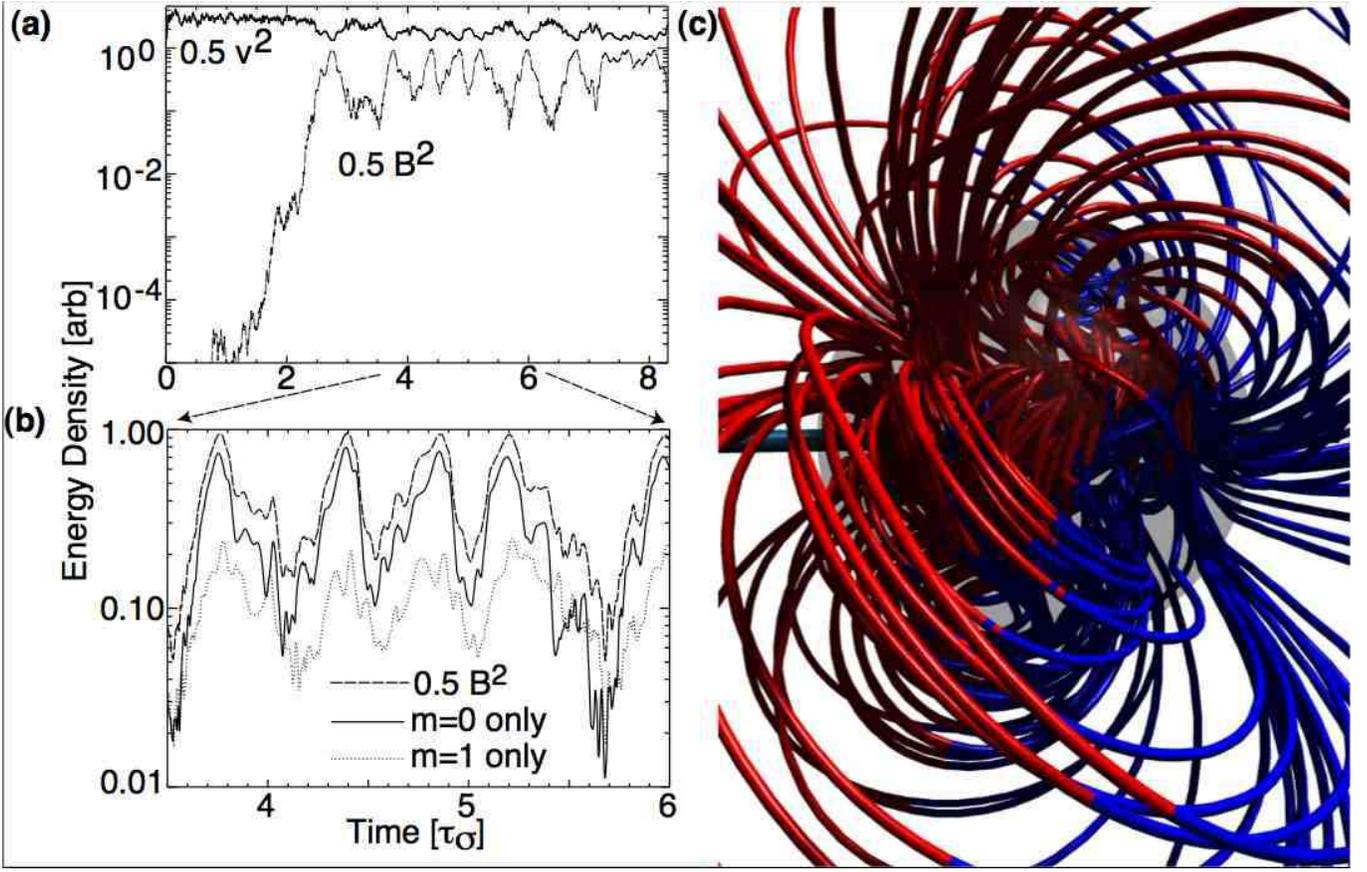}
\caption[Energy density of a turbulent dynamo and modal energy decomposition]
{(a) The energy density of a turbulent dynamo with $Rm=337$ and $Re=674$.  
(b) The energy density of the axisymmetric magnetic field ($m=0$) and 
the nonaxisymmetric dynamo ($m=1$). (c) Magnetic field lines of the turbulent saturated dynamo
\label{fig:9}}
\end{figure*}

\begin{figure}
\includegraphics[width=\figurewidthtwo]{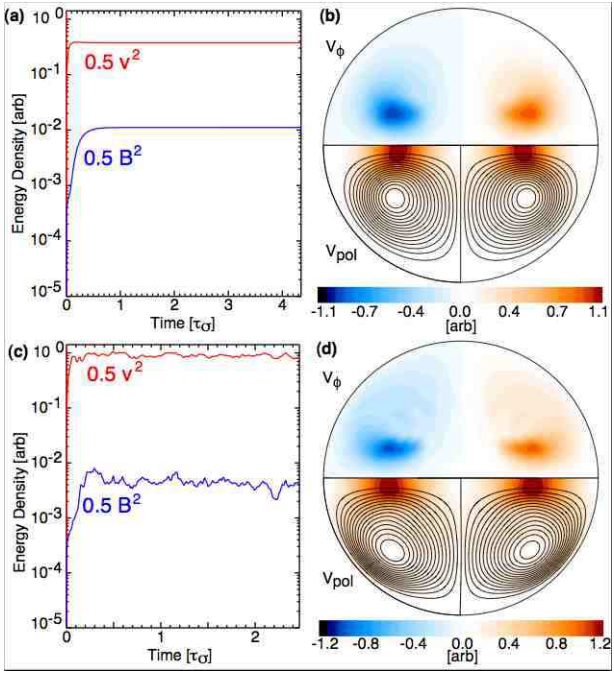}
\caption[Energy densities and mean flows from laminar and turbulent
applied-field simulations] {Simulations with an externally-applied,
axisymmetric magnetic field. (a) Kinetic and magnetic energy densities
for a Re=116 (laminar), Rm= 70 (subcritical) simulation. (b) the
resulting flow.  (c) Kinematic and magnetic energy densities for an
Re=1803 (turbulent), Rm= 108 (subcritical) simulation. (d) The
axisymmetric, time averaged velocity field for the turbulent simulation.
The time average is over the time interval 0.3-2.4
$\tau_{\sigma}$, which is roughly
$30$ decorrelation times.
\label{fig:10}}
\end{figure}

\begin{figure}
\includegraphics[width=\figurewidthtwo]{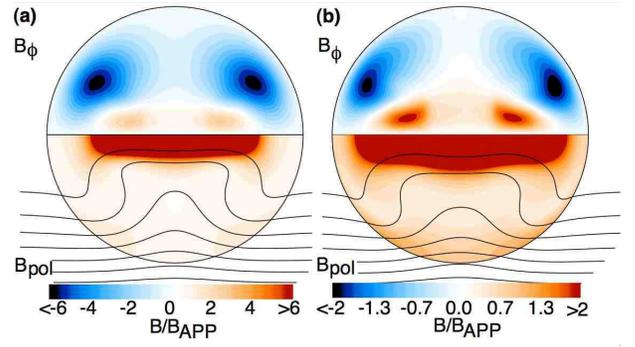}
\caption[Magnetic field and current of laminar flow and turbulent flows with an applied magnetic field]
{(a) The magnetic field for a laminar flow
described in Fig.\ \ref{fig:10}.  The resulting
total magnetic field (the sum of the externally applied field and those
generated by the currents in the liquid metal) is shown as a multiple of the
applied magnetic field.  
(b) The time-averaged magnetic fields, scaled to the applied field, for a turbulent
flow [Rm=107 (subcritical), Re=1803].  
The peak internal poloidal magnetic field is $9.3$ times
larger than the applied field.  
\label{fig:11}}
\end{figure}

\begin{figure}
\includegraphics[width=\figurewidthtwo]{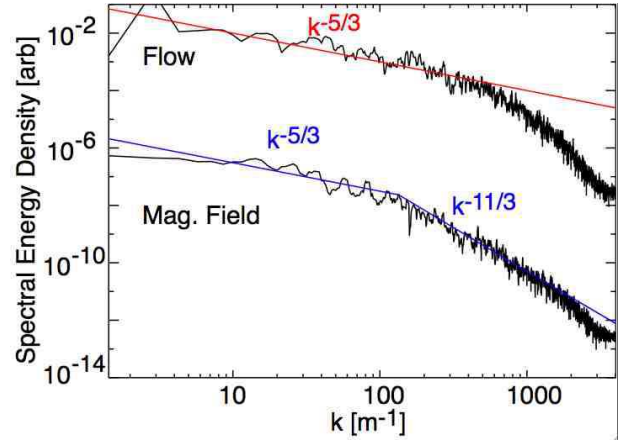}
\caption[Energy spectra of MHD turbulence]
{The wavenumber spectrum computed from frequency spectrum of
fluctuations from $6 \tau_{\sigma}$ of flow (fitted with the red
$k^{-5/3}$ curve) output at a position ($r\sim0.75 \ a$,
$\theta\sim\pi/2$, $\phi$=$0$) with a weak applied magnetic field of
$B_0\sim 51.3$ G, fitted with the blue $k^{-5/3}$ at low $k$ and
$k^{-11/3}$ at large $k$.  For this simulation, $Rm=130$, and $Re=1450$
and fluctuations are assumed to be due to convection of spatial
variations in the field. The dispersion relation is $\omega = k
\left<V\right>$. \label{fig:12}}
\end{figure}

\begin{figure}
\includegraphics[width=\figurewidthtwo]{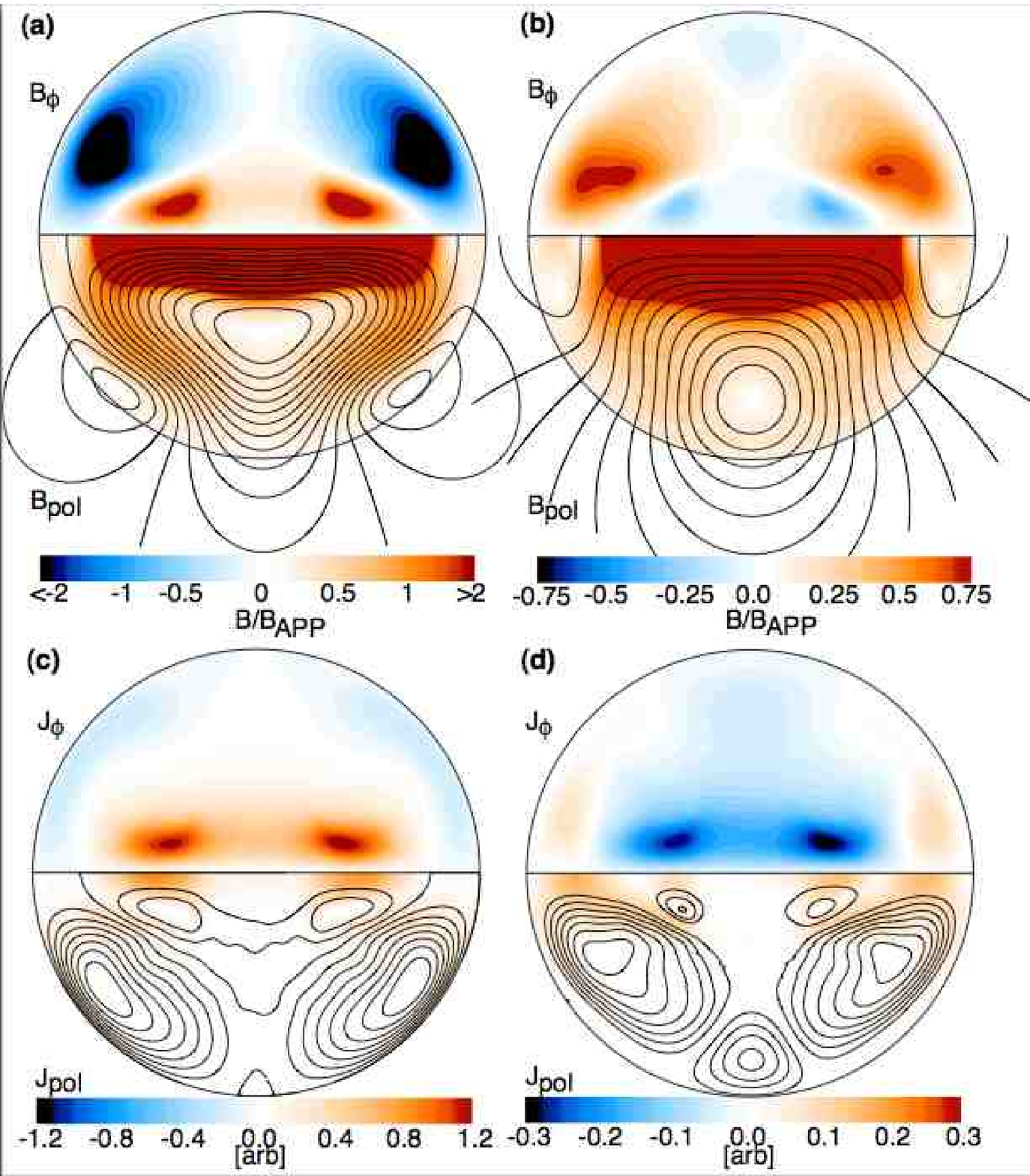}
\caption[EMF induced magnetic fields and currents]
{The magnetic field (a) 
generated by the mean-flow EMF $\expec{\bmr{V}}\times\expec{\bmr{B}}$. The
peak internal poloidal magnetic field is $11.2$ times larger than the
applied field. The magnetic field (b) 
generated by the turbulent EMF $\expec{\widetilde{\bmr{v}}\times\widetilde{\bmr{b}}}$.
(c) The currents associated with (a), and 
(d) the currents associated with (b).
The turbulent flow  has 
[$Rm=107$ (subcritical), $Re=1803$] and an externally generated magnetic field
applied along the symmetry axis.  The time average is taken over 
2.1$\tau_{\sigma}$.  Magnetic fields are scaled to the strength
of the applied field.}
\label{fig:13} 
\end{figure}

\begin{figure}
\includegraphics[width=\figurewidthtwo]{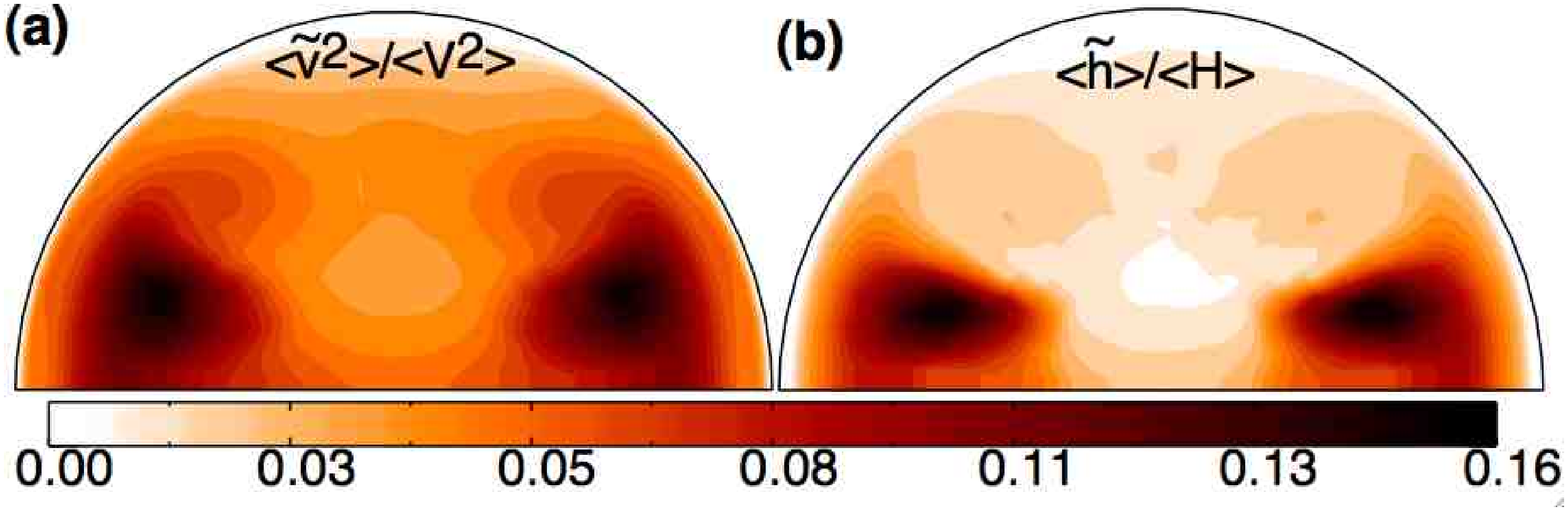}
\caption[spatial variation and rms kinetic helicity of MHD turbulence]
{The turbulent and helical fluctuations for the simulation described in
Fig.\ \ref{fig:12}
 (a) The average of the squared turbulent fluctuations
as a multiple of the peak squared mean flow. (b) The time average
of the kinetic helicity fluctuations $\expec{\widetilde{\bmr{v}}\cdot
  \nabla \times\widetilde{\bmr{v}}}$ as a multiple of the volume-averaged helicity of
the mean flow. \label{fig:14}}
\end{figure}

\section{Turbulent Dynamos \label{sec:turb_dynamos}}
To investigate the effect of turbulence on the dynamo transition,  
simulations are performed at lower $Pm$ (higher $Re$). 
The flow changes from laminar to turbulent at $Re\approx 420$.  Above
this value, a hydrodynamic instability grows exponentially on 
approximately an eddy-turnover time scale $\tau_c$ with a predominantly
 $m=1$ spatial structure. Through
nonlinear coupling, the instability quickly leads to
strongly-fluctuating turbulent flows (a detailed discussion of the
spectrum of the turbulence is  deffered to Sec.\ V).  
 Fluctuations about the mean flow
exist at all scales, including variations in the large-scale flow
responsible for the dynamo.  The turbulence is inhomogenous with
boundary layers, localized forcing regions, and strong shear layers.

The effect of these fluctuations on the dynamo onset conditions and on
the resulting saturation mechanism depends sensitively upon the
viscosity (parameterized by $Pm$).  Figure\
\ref{fig:5} shows an example of the broad range of dynamics
exhibited by decreasing $Pm$, for an approximately fixed value of $Rm$.
The magnetic field dynamics fall into several regimes depending on $Re$:
the laminar dynamo, a dynamo that starts turbulent, 
but relaminarizes the saturated flow, a
turbulent dynamo, and finally a turbulent flow with no dynamo.  At $Pm =
1$, the viscosity is large enough to keep both the magnetic field and
velocity field fully laminar. The spectrum is dominated by the driven
velocity field and by the magnetic eigenmode, and the saturation
mechanism is the Lorentz braking and modification to the flow mentioned
above.

For $Pm = 0.33$, Fig.\ \ref{fig:5} shows a flow that is
initially turbulent, but the saturated state is laminar.  
The turbulent saturation of the
magnetic field results in a reduction in the fluctuations of the flow
since the Lorentz braking has reduced $Re$ below the hydrodynamic
instability threshold (decreasing $Re$ from 496 to 320).  A 
hydrodynamic case, which evolves the flow with $\bmr{B}=0$,
 shows that flow turbulence persists without the
addition of a magnetic field into the system.  The $Re$ threshold
distinguishing the turbulent saturated state from a relaminarized
saturation is $Re\sim 630$.

If $Rm$ is fixed near the experimental maxima while $Re$ is increased
beyond $700$, no dynamo is observed.  Despite the fact that the mean flow
still satisfies the requirements of a kinematic dynamo, the turbulent
flow does not produce a growing magnetic field.  Evidently, 
it is the turbulent
fluctuations about the mean flow that prevent field growth.  Using the
mean flow (averaged over several resistive times) for the $Pm=0.22$
(with $Rm$=190, $Re$=863) as a prescribed flow in a kinematic evolution
of the induction equation gives $Rm_{crit}\approx 93$, as shown in Fig.\
\ref{fig:6}.
Even though the average flow has $Rm$ well above $Rm_{crit}$ there is no
dynamo.  However, when the conductivity is doubled such that $Rm =388$ a
turbulent dynamo reemerges in the simulation.  Hence, an empirical
critical magnetic Reynolds number, $Rm_{crit,T}$ can be defined which
depends on $Re$ through the degree of turbulent fluctuations in the
flow.

These results are consistent with
the dynamo transition being affected by
the turbulent resistivity
of Eq.\ \ref{sigma_t}.  From analysis of the simulation results,
the correlation time $\tau_c$, the eddy scale size $\ell_v$, i
and fluctuation levels 
$C=\tilde{v}/V_0$ have been determined in order to estimate the
parameters in $\sigma_{T}$ under the assumption that the homogeneous
turbulence results roughly apply to this bounded, inhomogeneous flow.
Typical volume averaged values measured in the $Rm=190, \ Re=863$
simulation are: $\ell_v=0.022 a$, $C=0.45$, and $\tau_{corr} =
0.041\tau_{\sigma}$, which yields a volume-averaged conductivity
reduction of $\sigma_T/\sigma = 0.461$. The diminished conductivity
yields $Rm_{crit,T} = 238$.  The results from all of the simulations are
summarized in Fig.\ \ref{fig:7} which shows that an increasing
$Rm$, at fixed $Re$ reestablishes field growth where turbulent
fluctuations had previously suppressed the dynamo.  The dashed line in
Fig.\ \ref{fig:7} shows that the correlation length and constant
$C$ increase with $Re$ and eventually asymptote when the conductivity is
effectively reduced by $70$ \%.

The simulated turbulence has no \textit{de facto} scale separation.
This might appear to pose a problem, given that our interpretation of
the effect of turbulence is the introduction of a turbulent resistivity,
and the turbulent resistivity of Mean Field Theory (MFT) \cite{krause80} 
is usually couched in scale
separation arguments.  However, it should be noted that the scale
separation requirement associated with the $\alpha$ and $\beta$ effects
of MFT does not enter into the form of $\beta$, but does guarantee that
$\alpha<\beta$.  This is because $\alpha$ is proportional to helicity
whereas $\beta$ is proportional to energy while $\alpha$ multiplies a
lower derivative of the mean field than does $\beta$.  In this sense the
lack of scale separation in the simulations is consistent with the
apparent weakness of a turbulent-$\alpha$ effect in a regime with a
turbulent resistivity.

While the simulations are limited to $Re<2000$ by computational speed 
and storage, 
we believe the simulations capture the dominant effect since
the fluctuations at the largest scales are the strongest contributers to
the turbulent resistivity by the following argument.  In Kolmogorov
turbulence \cite{k41} the spectrum is $E(k)\propto\epsilon^{2/3}k^{-5/3}$, where
$\epsilon$ is the energy dissipation rate.  Thus the turbulent
resistivity goes as $[\int_{k_0}^{k_\nu}q^{-2}E(q)dq]^{-1/2}\sim
\epsilon^{1/3}k_0^{-4/3}$, where $k_0$ is the wavenumber of the large
scale eddies and $k_{\nu}$ is the dissipation scale wavenumber.  
In K41 turbulence,
$k_{\nu}\propto Re^{3/4}$, as $Re$ becomes large in comparison to $Rm$,
the effect of turbulent fluctuations on conductivity will asymptote to a
fixed value. It should be noted that the simple dimensional analysis
used for estimating the turbulent resistivity reflects isotropic
homogenous turbulence and is derived in the limit that
there is no mean flow; this dynamo relies almost entirely on the presence
of a mean flow.

An alternative viewpoint, consistent with the phenomenological
interpretation of enhanced resistivity put forward here, is that the
large-scale variations in the velocity field are continuously changing
the spatial structure and growth rates of the magnetic eigenmodes of the
system.  A more thorough treatment of the dynamic variation of dominant
eigenmodes can be found for a slightly different problem in
\cite{terry06}.  Two effects can be important.  First, the instantaneous
growth rate of the least damped eigenmode fluctuates between growing and
damped. For a run with $Rm=193$, $Re=893$ and $\epsilon=0.4$, shown in
Fig.\ \ref{fig:8}, a dynamo occurs only when the flow spends sufficient
time in phases which are kinematic dynamos.  The kinematic growth rate
is most often positive, consistent with the time averaged flow having
growing magnetic field solutions, yet the modifications made to the flow
during the subcritical periods are sufficient to stop the dynamo.
Second, the turbulence couples energy from the growing magnetic
eigenmode into spatially-similar damped eigenmodes.  As the flows
evolve, the spatial structure of the eigenmodes change.  The magnetic
field structure of a single eigenmode at some previous instant in time
must be described in terms of several modes after the flow changes.
This transfer of energy from the primary mode is equivalent to enhanced
dissipation.  Analysis of the eigenmode structure shows that the least damped eigenmode
during a nondynamo phase in Fig.\ \ref{fig:8} varies between the marginally 
stable nonaxisymmetric dipole and a stable axisymmetric magnetic mode.  Thus the 
flow imparts energy to a spatially similar, but distinct magnetic eigenmode.

Finally, it should be noted that distinguishing
 between growing and damped magnetic fields is difficult
in the turbulent simulations.  Typically, the turbulent runs have been
limited to durations of less than 10 $\tau_\sigma$. 
The transition may also be considerably more complex
as seen  in Fig.\ \ref{fig:5} where
magnetic energy of the $Pm=0.22$ simulation 
may show intermittent growth near $Rm_{crit,T}$.
The simulations are thus consistent with intermittent
excitation of the dynamo eigenmode by the mean flow.  The peak magnetic
energy is limited by the magnitude of the initialized noise the simulation
is started with instead of the backreaction with the flow.  This effect
is especially relevant when the magnetic field is sustained by an
external source as shown in Section \ref{sec:applied}.

A dynamo still occurs in these flows for sufficiently
large $Rm$ (keeping $Re$ fixed, growing magnetic energy is detected for
sufficiently large $Pm$).  
An example of a time evolution and spatial structure of
a saturated turbulent magnetic field is shown in Fig.\ 
\ref{fig:9} for a simulation with $Rm=337$ and $Re=674$.
An $m=1$ transverse dipole field is still 
present, as in the case
of the laminar dynamo in Fig.\ \ref{fig:2}, however this turbulent
dynamo is now dominated by the presence of a large $m=0$ field, aligned with
the axis of symmetry of the impellors as shown in
Fig.\ \ref{fig:9}(b).
This component, by itself, would appear to violate 
Cowling's theorum \cite{cowling33}, and so it must result from
non-axisymmetric components of the velocity field and the magnetic field.
Thus it appears probable that the nature of dynamo has changed
fundamentally from a simple eigenmode driven by the two vortex flow, to a 
dynamo in which the turbulent fluctuations may be responsible for
generating the large-scale magnetic field.

\section{Simulations of a subcritical turbulent flow with a weak,
externally-applied magnetic field\label{sec:applied}}
As a means of further highlighting the different physics and conditions
underlying turbulent and laminar dynamos, subcritical flows are
simulated with focus on the potential role of fluctuation driven
currents in the self-excitation process.  Subcritical flows have
$Rm<Rm_{crit}$, and are not expected to lead to self-excited magnetic
fields.  The MHD behavior is investigated by applying a magnetic field
which is generated by currents flowing in coils external to the sphere.
The configuration studied is similar to the set of experiments described
in Ref.\ \cite{spence06}, 
and is deliberately set up as an axisymmetric
system in which fluctuation driven currents can be easily detected.

The numerical technique employed is similar to the dynamo simulations
described above in all but one respect, namely a different boundary
condition is used with $\lm{C}\ne0$ in Eq.\ \ref{eq:Bpbc}.
These boundary conditions match the magnetic field to a scalar magnetic
potential $\bmr{B}=-\nabla\Phi_m$, which is valid
in the region between the surface of the sphere and
the external magnets.  $\Phi$ satisfies Laplace's equation
and its solution is well known:
\begin{equation}
\Phi_m(r,\theta,\phi)=\sum_{\ell,m}\left( 
A_{\ell,m} r^{\ell} + D_{\ell,m}r^{-(\ell+1)}\right)
Y_\ell^m(\theta,\phi),
\end{equation}
where $Y_\ell^m(\theta,\phi)$'s are the spherical harmonics. The
$D_{\ell,m}$ terms represent the magnetic field generated by currents in
the sphere, and the coefficients $A_{\ell,m}$ can be chosen to describe
a magnetic field of arbitrary shape and orientation applied by currents
external to the sphere.  In this paper and
in the simulations described below, a uniform magnetic field is applied
along the symmetry axis of the forcing terms, and is characterized by a
single coefficient $A_{1,0}$, all higher order terms being zero.  The
applied magnetic field, $B_{1,0}$, is weak enough so that it does not
alter the large-scale flow.
The strength of the applied magnetic field
is moderated by keeping the Stuart number $N\equiv \sigma a
B_{1,0}^2/\rho \bmr{v}_0<0.1$. In sodium, with a $Rm\sim100$, $N\sim0.1$
would correspond to an applied field of $156$ gauss.  The applied field
for these simulations is uniform and applied along the impeller axis with
 $B_{1,0}\sim57$ gauss and $N\sim 10^{-2}$.
However, since the velocity fluctuations decrease with scale, 
the Stuart number increases with scale, indicating the Lorentz force may influence
small-scale fluid motion.  Examples of such simulations are shown in
Fig.\ \ref{fig:10}, where the kinetic energy and magnetic
energy are shown for laminar and turbulent runs.

For the laminar flow, the induced currents and resulting magnetic field
are purely due to the magnetic field interacting with the mean flow, as seen
in Fig.\ \ref{fig:11}(a).  Two main effects are observed.  First,
induced toroidal currents compress lines of poloidal magnetic field near
the axis of the device.  The lines are pulled outward at the poles and
inward at the equator.  The net result is a reduction of the poloidal
field strength at the equator in the outer region, and a large
amplification at the axis (the peak poloidal field is $18$ times the
applied field).  Second, poloidal currents generate a toroidal magnetic
field.  These currents are generated by the well-known omega effect of
dynamo theory whereby differential toroidal rotation of the fluid is
able to stretch the field into the toroidal direction \cite{moffatt78}.
The amplitude of the peak toroidal field is greater than $6$ times the
applied field.

The transition to turbulence is still characterized by the same
$Re\sim420$ threshold described above, since the Stuart number for the
applied magnetic field is small.  Below this threshold, the
nonaxisymmetric part of the flows is negligible while above this
threshold nonaxisymmetric fluctuations in both $\bmr{B}$ and $\bmr{V}$
can be as large as $40\%$ of the mean values.  The geometry of the
simulations (axisymmetric drive terms aligned with the applied magnetic
field) makes it possible to separate mean, axisymmetric quantities and
fluctuating quantities,
\begin{equation}
\bmr{B}=\expec{\bmr{B}}+\widetilde{\bmr{b}} \mbox{ and }
\bmr{v}=\expec{\bmr{V}}+\widetilde{ \bmr{v}},  
\end{equation}
where the brackets denote a time average over several resistive times.
In practice, $\expec{\bmr{B}}$ and $\expec{\bmr{V}}$ are axisymmetric
for sufficiently long time averages.  Using these definitions, the
time-averaged magnetic fields can be computed for laminar
and turbulent flows, shown in Fig.\ \ref{fig:11}.

Both laminar and turbulent flows demonstrate toroidal field production
and expulsion of poloidal flux.  Laminar and turbulent results differ in
several important ways, however, which are attributable to the currents
being driven by MHD fluctuations.  First, the toroidal field is greatly
reduced in the turbulent run.  The induced toroidal field is $6$ times
the applied field strength in the laminar flow and is only twice the
applied field in the turbulent case.  Second, the 
peak poloidal field is halved
in the turbulent run, as shown in Fig.\ \ref{fig:11}(a).
Third, there is a net magnetic dipole moment associated with the
induced field which is not present in the laminar case.  These
differences are partially the 
result of a difference between the mean flows in the two cases,
but are mostly a due to a strong influence of the turbulence on the
current generation.
This can be interpreted in the context of a modification to 
the mean-field Ohm's law, i.e.\ turbulence-generated currents are
modifying the large-scale, mean magnetic field.

A turbulent EMF is possible because of the
flow fluctuations and the magnetic field generated  by 
the passive advection of the applied magnetic field by
the Kolmogorov-like turbulence in the velocity field.  Fig.\
\ref{fig:12} shows the wavenumber spectrum as estimated from the
frequency spectrum of the fluctuations in both $V$ and $B$
at a fixed point in the
simulation using the Taylor hypothesis to map frequency fluctuations to
wavenumber $\omega\sim k \left<V\right>$. 
It is clear that both the velocity field and magnetic field
have an inertial range ($k^{-5/3}$) and a dissipation scale, although
the dissipation scales are at different values of $k$.  The $k^{-5/3}$
scaling of the velocity field (the inertial range) is expected from the
Kolmogorov theory of hydrodynamic turbulence.  The dissipation scale for
the fluid turbulence is expected to be at $k_{\nu}\sim Re^{-3/4}=235$
which is roughly the position of the viscous cutoff shown in Fig.\
\ref{fig:12}.  The limited inertial range at low k, is primarily
due to constraints on long-time averages of the data imposed by
computational speed.

The $k^{-5/3}$ scaling of the magnetic field corresponds to the
weak-field approximation in which the induced magnetic fluctuations are
due to advection of the mean magnetic field by the mean flow
\cite{nornberg06} for $k<k_{\sigma}\sim Rm/a$.  The
$k^{-11/3}$ power law results from a balance between the mean magnetic
field advected by turbulence and the resistive dissipation of magnetic
fluctuations.
The dissipation scales are evident from the knee in the wave number
spectra of Fig.\,\ref{fig:12}. The spectrum is constructed from
the power spectrum of the value of $B_r$ near the equator.
Consequently, the magnetic field gains structure at smaller scales as
$Rm$ increases, down to scale sizes of $\ell_\sigma = 2\pi/k_\sigma =
4.8$\,cm at $Rm=130$.

The simultaneous fluctuating magnetic and velocity fields can
potentially drive current in a mean-field sense.  The motional EMF can
be written as
\begin{equation}
\bmr{v}\times\bmr{B}=
\left< \bmr{V} \right> \times \left< \bmr{B} \right> +
\left< \bmr{V} \right> \times  \widetilde{\bmr{b}}  +
\widetilde{\bmr{v}} \times \left< \bmr{B} \right> +
\widetilde{\bmr{v}}\times\widetilde{\bmr{b}},
\end{equation}
where the mean-fields have been separated from the fluctuating parts.
The time averages must be taken over times long compared to a turbulent
decorrelation time and comparable to the resistive diffusion time.
Since the turbulent decorrelation time, $\tau_C\sim0.05\tau_{\sigma}$,
integrating the induction term over several
resistive times yields
\begin{equation}
\label{eq:time_var_ind}
\expec{\bmr{v}\times\bmr{B}}=\expec{\bmr{V}}\times\expec{\bmr{B}}+
\expec{\widetilde{\bmr{v}}\times\widetilde{\bmr{b}}}.
\end{equation}
An important question is whether the currents generated in the
simulation are primarily 
due to the motional EMF associated with
the mean-flow and the mean magnetic field,
$\expec{\bmr{V}}\times\expec{\bmr{B}}$, or if there are also 
currents driven by the turbulent EMF
$\expec{\widetilde{\bmr{v}}\times\widetilde{\bmr{b}}}$.  This can be
investigated by examining the various terms in Ohm's Law
\begin{equation}
\label{eq:nd_ohms}
\bmr{E}=\eta \bmr{J}-
\expec{\bmr{V}}\times\expec{\bmr{B}}+
\expec{\widetilde{\bmr{v}}\times\widetilde{\bmr{b}}}
\end{equation}
It is clear that in steady-state there can be no inductive electric
field in the toroidal direction since the poloidal flux is constant.
Axisymmetry precludes an electrostatic potential from driving current in
the toroidal direction, and so the toroidal current can only be
generated by the mean-flow and the turbulent EMF.  Thus any currents
driven in the toroidal direction contribute to the poloidal magnetic
field.  Fig.\ \ref{fig:13}(c) shows the currents driven by these
fluctuations and their corresponding magnetic field (a).  The
fluctuation-induced magnetic field is $3.5$ times larger than the
applied field and comprises a third of the total field strength.

It has been recently shown that an axisymmetric flow and axial
magnetic field cannot induce a dipole moment in any simply-connected
bounded system \cite{spence06}.  This is essentially due to the fact that
the flow outside the conducting region is zero, while the streamlines of
flow perpendicular to the magnetic flux are closed and bounded within
the conducting region.  Only a turbulent EMF can create the dipole
moment.  With a weak applied field in a turbulent fluid, averaging over
several eddy turnover times and averaging along $\bhat{\phi}$ eliminates
the nonaxisymmetric component of the current, therefore the only
nontrivial component of the dipole moment is
\begin{equation}
\mu_Z=\oint d^3x r \sin{\theta} J_{\phi}.
\end{equation}
The toroidal current generated by
$\expec{\widetilde{\bmr{v}}\times\widetilde{\bmr{b}}}$ 
from Eq.\ \ref{eq:time_var_ind}, is shown
in Fig.\ \ref{fig:13}(d) and the associated dipole moment (antiparallel to
the external field) is clearly seen in Fig.\ \ref{fig:13}(b).  Alternatively, 
the EMF due to $\bmr{V}$ and $\bmr{B}$ gives
rise to a hexapole magnetic field (in Fig.\ \ref{fig:13}(a)).  
The resulting poloidal field
reduces the surface magnetic field by $20\%$.  The largest values of the
turbulent toroidal current occur where the omega effect is also large.

The EMF which generates the toroidal current associated with the
induced dipole moment may very well resemble the currents driven by
the  well-known $\alpha$-effect.  The omega effect generates a
toroidal magnetic field  which would in turn would support
a current of the form $J_\phi = \alpha B_\phi$.   
It is impossible to uniquely identify the
current this way, however.  A non-uniform $\beta$-effect (change in
local resistivity) could equally well explain the results. To do this
would require separating the currents
associated with the helical fluctuations from the non-helical
fluctuations and this has not yet been done.  A local analysis of the
turbulent helicity content in Fig.\ \ref{fig:14}(b) shows that helical 
fluctuations exist that might be expected to drive a current through the
$\alpha$-effect.

To study Ohm's law in the poloidal direction
requires  a full
treatment of the poloidal electric field since an electrostatic
potential is not ruled out by symmetry arguments.
In MHD, the electrostatic potential is assumed to instantaneously adjust
itself to ensure that $\bnabla\cdot \bmr{J}=0$. This can only be assured
if the divergence of the motional EMFs is balanced by a spatially
varying electric field
\begin{equation}
\label{eq:divE}
\divergence{\bmr{E}}=-\laplacian{\Phi}-\laplacian{\tilde{\Phi}}=
\divergence{\bigl(\expec{\bmr{V}}\times\expec{\bmr{B}}+
\expec{\widetilde{\bmr{v}}\times\widetilde{\bmr{b}}}\bigr)},
\end{equation}
where $\Phi$ and $\tilde{\Phi}$ are electrostatic potentials due to the
stationary EMF, and turbulent EMF respectively.  
Thus, a poloidal current can be associated with the mean-flows and the
turbulent EMF respectively:
\begin{eqnarray}
\bmr{J}_{pol} &  =  & \sigma \left( -\nabla \Phi + \expec{\bmr{V}}
\times\expec{\bmr{B}}_{pol} \right) \\
\tilde{\bmr{J}}_{pol} &  =  & \sigma \left( -\nabla \tilde\Phi +
 \expec{\widetilde{\bmr{v} }\times\widetilde{ \bmr{b} } } \right).
\end{eqnarray}
When analyzing Ohm's law in the poloidal direction, it is necessary to
first compute these potentials, which has been done for the poloidal
currents in Fig.\ \ref{fig:13}.

The simulations indicate there is a strong poloidal current, shown in
Fig.\ \ref{fig:13}(d), associated with the fluctuations.  The current
acts to greatly reduce the toroidal magnetic field generated by a
comparable laminar flow, thereby reducing the toroidal field in the
core.  This resembles the diamagnetic $\gamma$-effect \cite{krause80},
due to gradients in the turbulence intensity.  The $\alpha$-effect is
the diagonal part of a mean-field tensor: $\bmr{J} = \sigma \bm{\alpha}
\cdot \bmr{B}$.  The off-diagonal terms can also be written so that
$\bmr{J} = \sigma \bm{\gamma} \times \bmr{B}$.  Figure\ \ref{fig:14}(a)
shows the squared velocity fluctuations decrease away from the axis of
symmetry with the polar radius, $\rho$.  For isotropic turbulence, the
inhomogeneity in the fluctuations would give rise to a $\gamma$-effect
of the form $-\sigma \bm{\gamma}(\rho)\times \bmr{B_{\Tor}}$ with
$\bm{\gamma}\propto \bnabla \mathrm{v}^2$.  The poloidal current due to
turbulent diamagnetism would counteract the toroidal magnetic field
caused by the omega effect.  Comparison between Fig.\ref{fig:13}(a)
with Fig.\ref{fig:14}(b) shows that regions of steep gradients in the
turbulent fluctuations correspond to regions of strong fluctuation
induced poloidal current.

\section{Summary  \label{sec:summary}}

The role of turbulence in generating current and moderating the growth
of magnetic fields has been studied for the Madison Dynamo experiment
using 3D numerical simulations.
A simple forcing term has been used to model impellors in the
experiment; at sufficient forcing the flow becomes turbulent.
Two regimes were explored: one with an external applied
magnetic field and flow subcritical to the dynamo instability and one
with no external field and super-critical flow.  The role of the
turbulence on current generation and self-excitation is marked.

The onset conditions for the dynamo instability are governed not only by
$Rm$ but also by the
magnetic Prandtl number $Pm$.  At $Pm\sim1$, the transition and
saturation agree with laminar predictions and are considered laminar
dynamos.  At lower $Pm$, $\rmc$ increases, consistent with a reduction in
conductivity due to turbulent fluctuations. However, at higher $Rm$ the
character of the dynamo changes; its symmetry suggests that turbulence
driven 
currents are important in the self-excitation process.
The $Pm$ values in the
simulations are still orders of magnitude larger than in liquid-metal
experiments (and for geo and solar dynamos) due to memory and speed
limitations of computers, and so experiment support is critical
for verifying these results.

To quantify currents driven by fluctuations in the experiment,
simulations of subcritical flows have been performed, and the currents
driven by the turbulent fluctuations have been observed directly.  The
main effect of the turbulence on an externally-applied magnetic field is
the reduction of field strength compared with those computed for laminar
flows.  The laminar two-vortex flow compresses the applied poloidal
magnetic flux near the axis of symmetry and through toroidal flow shear
creates a strong toroidal magnetic field.  Both effects are reduced in
turbulent flows.  The mean flow produced at large Reynolds numbers
differs from its laminar counterpart, which accounts for some of the
discrepancy between the build-up of toroidal field and flux compression
of the poloidal field observed in the laminar and turbulent fluids.
However, it has also been shown that a fluctuation driven EMF drives
current which modifies the large-scale magnetic field, both generating a
dipole moment and expelling toroidal flux from the interior region.

We would like to thank C. Sovinec, S. Prager, J. Wright, and E. Zweibel
for many useful discussions.  This work was supported by the National
Science Foundation.


\begin{thebibliography}{10}

\bibitem{moffatt78}
H. Moffatt, {\em Magnetic field generation in electrically conducting fluids}
  (Cambridge University Press, Cambridge, 1978).

\bibitem{childress95}
S. Childress and A. Gilbert, {\em Stretch, Twist, Fold: The fast dynamo}
  (Springer, Berlin, 1995).

\bibitem{kraichnan67}
R. Kraichnan and S. Nagarajan, Phys. Fluid {\bf 10},  859  (1967).

\bibitem{frisch75}
U. Frisch, A. Pouquet, J. L\'{e}orat, and A. Mazure, J. Fluid Mech. {\bf 68},
  769  (1975).

\bibitem{pouquet78}
A. Pouquet and G. Patterson, J. Fluid Mech. {\bf 85},  305  (1978).

\bibitem{alexakis05}
A. Alexakis, P. Mininni, and A. Pouquet, On the inverse cascade of magnetic
  helicity, \url{http://arxiv.org/abs/physics/0509069}, 2005.

\bibitem{bullard54}
E. Bullard and H. Gellman, Phil. Trans. R. Soc. Lond. A {\bf 247},  213
  (1954).

\bibitem{roberts72}
P.~H. Roberts, Phil. Trans. Roy. Soc. London A {\bf 272},  60  (1972).

\bibitem{gubbins73}
D. Gubbins, Phil. Trans. R. Soc. Lond. A {\bf 274},  493  (1973).

\bibitem{glatzmaier84}
G. Glatzmaier, J. Comp. Phys. {\bf 55},  481  (1984).

\bibitem{glatzmaier95}
G. Glatzmaier and P. Roberts, Phys. Earth Plan. Int. {\bf 91},  63  (1995).

\bibitem{glatzmaier02}
G. Glatzmaier, Ann. Rev. Earth Planet. Sci.  in press  (2002).

\bibitem{kageyama95}
A. Kageyama and T. Sato, Phys. Plasmas {\bf 2},  1421  (1995).

\bibitem{kageyama97}
A. Kageyama and T. Sato, Phys. Plasmas {\bf 4},  1569  (1997).

\bibitem{kuang97}
W. Kuang and J. Bloxham, Nature {\bf 389},  371  (1997).

\bibitem{meneguzzi81}
M. Meneguzzi, U. Frisch, and A. Pouquet, Phys. Rev. Lett. {\bf 47},  1060
  (1981).

\bibitem{cattaneo96}
F. Cattaneo and D. Hughes, Phys. Rev. E {\bf 54},  4532  (1996).

\bibitem{mueller03}
W. M\"{u}ller, D. Biskamp, and R. Grappin, Phys. Rev. E {\bf 67},  066302
  (2003).

\bibitem{boldyrev04}
S. Boldyrev and F. Cattaneo, Phys. Rev. Lett. {\bf 92},  144501  (2004).

\bibitem{schekochihin04}
A. Schekochihin, S. Cowley, J. Maron, and J. McWilliams, Phys. Rev. Lett. {\bf
  92},  054502  (2004).

\bibitem{ponty05}
Y. Ponty, P. Mininni, D. Montgomery, J. Pinton, H. Politano, and A. Pouquet,
  Phys. Rev. Lett. {\bf 94},  164502  (2005).

\bibitem{gailitis00}
A. Gailitis, A. Lielausis, E. Platacis, S. Dement'ev, A. Cifersons, G. Gerbeth,
  T. Gundrum, F. Stefani, M. Christen, H. H\"{a}nel, and G. Will, Phys. Rev.
  Lett. {\bf 84},  4365  (2000).

\bibitem{gailitis01}
A. Gailitis, A. Lielausis, E. Platacis, S. Dement'ev, A. Cifersons, G. Gerbeth,
  T. Gundrum, F. Stefani, M. Christen, and G. Will, Phys. Rev. Lett. {\bf 86},
  3024  (2001).

\bibitem{gailitis04}
A. Gailitis, A. Lielausis, E. Platacis, G. Gerbeth, and F. Stefani, Phys.
  Plasmas {\bf 71},  2838  (2004).

\bibitem{stieglitz01}
R. Stieglitz and U. M\"{u}ller, Phys. Fluids {\bf 13},  561  (2001).

\bibitem{mueller04}
U. M\"{u}ller and R. Stieglitz, Phys. Fluids {\bf 16},    (2004).

\bibitem{mueller04a}
U. M\"{u}ller and R. Stieglitz, J. Fluid Mech. {\bf 498},  31  (2004).

\bibitem{reighard01}
A. Reighard and M. Brown, Phys. Rev. Lett. {\bf 86},  2794  (2001).

\bibitem{mueller02}
U. M\"{u}ller and R. Stieglitz, Nonlinear Processes in Geophysics {\bf 9},  165
   (2002).

\bibitem{roberts00}
P. Roberts and G. Glatzmaier, Rev. Mod. Phys. {\bf 72},  1081  (2000).

\bibitem{cattaneo02}
F. Cattaneo,  in {\em Modeling of Stellar Atmospheres}, {\em ASP Conference
  Series}, edited by N. Piskunov, W. Weiss, and D. Gray (IAU Publications,
  Paris, 2002).

\bibitem{petrelis03}
F. P\'etr\'elis, M. Bourgoin, L. Mari\'e, J. Burguete, A. Chiffaudel, F.
  Daviaud, S. Fauve, P. Odier, and J.-F. Pinton, Phys. Rev. Lett. {\bf 90},
  174501  (2003).

\bibitem{peffley00}
N. Peffley, A. Cawthorne, and D. Lathrop, Phys. Rev. E {\bf 61},  5287  (2000).

\bibitem{oconnell00}
R. O'Connell, R. Kendrick, M. Nornberg, E. Spence, A. Bayliss, and C. Forest,
  in {\em Proceedings of the NATO Advanced Research Workshop on Dynamo and
  Dynamics, a Mathematical Challenge, Carg\'ese, France}, Vol.~26 of {\em Nato
  Science Series}, edited by P. Chossat, D. Armbustier, and I. Oprea (Kluwer,
  Dordrecht, 2001), p.\ 59.

\bibitem{quartapelle95}
L. Quartapelle and M. Verri, Computer Physics Communications {\bf 90},  1
  (1995).

\bibitem{canuto88}
C. Canuto, M. Hussaini, A. Quarteroni, and T. Zang, {\em Spectral Methods in
  Fluid Dynamics} (Springer-Verlag, Berlin, 1988), p.\ 84.

\bibitem{dudley89}
M. Dudley and R. James, Proc. R. Soc. Lond. A {\bf 425},  407  (1989).

\bibitem{holme96}
R. Holme and J. Bloxham, J. Geophys. Res. {\bf 101},  2177  (1996).

\bibitem{nornberg06}
M. Nornberg, E. Spence, R. Kendrick, C. Jacobson, and C. Forest, 
Phys. Plasmas {\bf 13},  055901  (2006).

\bibitem{k41}
A.~N. Kolmogorov, Dokl. Akad. Nauk. {\bf 30},  299-303  (1941).

\bibitem{terry06}
P. Terry, D. Baver, and S. Gupta, Phys. Plasmas {\bf 13},  022307  (2006).

\bibitem{cowling33}
T. Cowling, Mon. Not. R. Astr. Soc. {\bf 94},  39  (1933).

\bibitem{spence06}
E. Spence, M. Nornberg, C. Jacobson, R. Kendrick, and C. Forest, Phys.
  Rev. Lett. {\bf 96 } 055002 (2006).

\bibitem{krause80}
K. Krause and K. R\"adler, {\em Mean--field Magnetohydrodynamics and Dynamo
  Theory} (Pergamon Press, Oxford, 1980).
 
\end{thebibliography}
\end{document}